\RequirePackage{fix-cm}
\documentclass[twocolumn,epjc3]{svjour3}
\smartqed  
\RequirePackage{graphicx}
\usepackage{lmodern}
\usepackage{amsmath}
\newcommand{\be}{\begin{equation}}
\newcommand{\ee}{\end{equation}}
\newcommand{\bn}{\begin{eqnarray}}
\newcommand{\en}{\end{eqnarray}}
\newcommand{\bes}{\begin{subequations}}
\newcommand{\ees}{\end{subequations}}

\usepackage{epstopdf}
\usepackage{widetext}
\usepackage{amssymb}
\journalname{Eur. Phys. J. C}
\begin{document}

\title{The simplest non-minimal matter-geometry coupling in the $f(R,T)$ cosmology}


\author{P.H.R.S. Moraes$^{1,a}$
\and
        P.K. Sahoo$^{2,b}$
}

\thankstext{1}{e-mail: moraes.phrs@gmail.com}
\thankstext{2}{e-mail: pksahoo@hyderabad.bits-pilani.ac.in}

\institute{ITA - Instituto Tecnol\'ogico de Aeron\'autica - Departamento de F\'isica, 12228-900, S\~ao Jos\'e dos Campos, S\~ao Paulo, Brazil
\and
Department of Mathematics, Birla Institute of Technology and Science-Pilani, Hyderabad Campus, Hyderabad-500078, India
}

\date{Received: date / Accepted: date}

\maketitle

\begin{abstract}

The $f(R,T)$ gravity is an extended theory of gravity in which the gravitational action contains general terms of both the Ricci scalar $R$ and trace of the energy-momentum tensor $T$. In this way, $f(R,T)$ models are capable of describing a non-minimal coupling between geometry (through terms in $R$) and matter (through terms in $T$). In this article we construct a cosmological model from the simplest non-minimal matter-geometry coupling within the $f(R,T)$ gravity formalism, by means of an effective energy-momentum tensor, given by the sum of the usual matter energy-momentum tensor with a dark energy contribution, with the latter coming from the matter-geometry coupling terms. We apply the energy conditions to our solutions in order to obtain a range of values for the free parameters of the model which yield a healthy and well-behaved scenario. For some values of the free parameters which are submissive to the energy conditions application, it is possible to predict a transition from a decelerated period of the expansion of the universe to a period of acceleration (dark energy era). We also propose further applications of this particular case of the $f(R,T)$ formalism in order to check its reliability in other fields, rather than cosmology. 

\end{abstract}

\section{Introduction}
\label{sec:int}

Einstein's General Theory of Relativity (GR) has connected the matter content of the Universe with the geometry of the fabric of the space-time, through the well-known field equations $G_{\mu\nu}=8\pi T_{\mu\nu}$ (in natural units, which shall be adopted in this article). The {\it lhs} of the above equation is the Einstein's tensor, which satisfies the Bianchi identities $\nabla_\nu G_\mu^{\nu}\equiv0$, while the {\it rhs} is the energy-momentum tensor, which for a perfect fluid, that is going to be assumed here, is characterized by three quantities: $4-$velocity $u^{\mu}$, proper density $\rho$ and pressure $p$. From the  Bianchi identities, the covariance derivative $\nabla_\mu$ of the energy-momentum tensor is null ($\nabla_\mu T_\mu^{\nu}=0$), which implies the conservation of matter throughout the universe evolution. The Einstein's field equations can be seen as constraints on the simultaneous choice of the metric $g_{\mu\nu}$ (which is contained in $G_{\mu\nu}$) and $T_{\mu\nu}$.

Even though geometry and matter are on the same footing, GR does not consider any possible effects of a non-minimal coupling between them.

That is not the case, e.g., for the recently elaborated $f(R,L_m)$ \cite{harko/2010} and $f(R,T)$ theories \cite{harko/2011}, for which $R$ is the curvature scalar, $f(R,L_m)$ is a function of $R$ and matter lagrangian density $L_m$, and $f(R,T)$ is a function of $R$ and trace of the energy-momentum tensor $T$. These theories predict a non-conservation of the energy-momentum tensor ($\nabla_\nu T_\mu^{\nu}\neq0$). If one drops the conservation of $T_{\mu\nu}$, the continuity equation does not hold any longer and the models predict the creation of matter as shown in \cite{harko/2015b,harko/2014}.

In fact, in $f(R,T)$ gravity theory, the $T-$dependence is motivated by the consideration of quantum effects and it is well known that quantum field theory in curved space-time yields the possibility of particle production \cite{parker/1968,parker/1971}. Such a possibility in both quantum theory of gravity and extended gravity theories with matter-geometry coupling may be a clue that there is a connection between these two \cite{liu/2016}. 

The non-conservative cosmological evolution has been considerably investigated nowadays. For instance, T. Josset et al. have considered dark energy effects as a consequence of energy conservation violation \cite{josset/2017}. In this way, the cosmic acceleration \cite{riess/1998,perlmutter/1999} itself could be an observable consequence of energy conservation violation. Such an approach was made later within the $f(R,T)$ gravity context \cite{shabani/2017}.

Despite its recent elaboration, the $f(R,T)$ gravity already presents a large number of applications \cite{shamir/2015,singh/2014,jamil/2012,cm/2016,mmm/2016,mc/2016,mrc/2016,moraes/2016,moraes/2015,moraes/2014,amam/2016,ms/2016}.

Particularly, some other relevant results obtained from $f(R,T)$ applications can be seen in the following references. In \cite{mam/2016}, the hydrostatic equilibrium equation (also referred to as Tolman-Oppenheimer-Volkoff equation) was constructed and numerically solved for neutron and quark stars. It has been shown that the term proportional to $T$ in the formalism yields an increment on the mass of these objects, making possible to predict the existence of massive pulsars recently detected \cite{demorest/2010,antoniadis/2013}.

A set of solutions describing the interior of compact stars under $f(R,T)$ gravity was generated in \cite{das/2016}. The acceptability of the model within observational constraints has been checked. Gravastars have been recently described in $f(R,T)$ theory in Ref.\cite{das/2017}. 

R. Zaregonbadi et al. \cite{zaregonbadi/2016} showed that the term in the field equations coming from $f(R,T)$ gravity leads to a flat rotation curve in the halo of galaxies, putting the dark matter paradigm in check. Moreover, solar system consequences of the $f(R,T)$ gravity models were investigated in \cite{shabani/2014}.

The above $f(R,T)$ gravity bibliography did not explore the consequences of a non-minimal matter-geometry coupling. In other words, there was no product between $R$ and $T$, or functions of them, in any of the functional forms worked in those references. It is the purpose of the present article to explore the cosmological consequences of a non-minimal matter-geometry coupling in $f(R,T)$ gravity. In order to do so, we will take the simplest coupling, such that $f(R,T)=f_1(R)+f_2(R)f_3(T)$, with $f_1(R)=f_2(R)=R$ and $f_3(T)=\alpha T$, with $\alpha$ a constant. 

Matter-geometry coupling subject has been deeply investigated. For instance, A. Connes showed that the foundation of non-commutative geometry could be related to a coupling between matter and geometry \cite{connes/1996}. New insights on matter-geometry coupling paradigm were presented in \cite{delsate/2012}. Modified gravity with arbitrary matter-geometry coupling was formulated within metric and Palatini formalism respectively in \cite{harko/2008,harko/2011b}. Furthermore, a thermodynamic interpretation of gravitational models with matter-geometry coupling was given in \cite{harko/2014}.

The present article is organized as follows: in Section 2 we present the basic mathematical formalism of the $f(R,T)=f_1(R)+f_2(R)f_3(T)$ and derive the field equations of the case $f_1(R)=f_2(R)=R$ and $f_3(T)=\alpha T$, which will be assumed here. We write our equations in terms of an effective energy-momentum tensor $T_{\mu\nu}^{eff}$, which is given by the sum of the usual matter energy-momentum tensor $T_{\mu\nu}$ and the dark energy term $T_{\mu\nu}^{DE}$, coming from the matter-geometry coupling predicted in the Theory. In Section 3, the Friedmann-like equations for such a model are constructed and the solutions for cosmological parameters such as scale factor, Hubble parameter and deceleration parameter are presented. We also plot the solution for the deceleration parameter in both time $t$ and redshift $z$. In Section 4 we apply the energy conditions in our solutions. The energy conditions tell us the range of values of the free parameters of the model which generate well-behaved cosmological scenarios. We, then, construct graphics of the quantities $\rho^{eff}$, $p^{eff}$, $\omega^{eff}=p^{eff}/\rho^{eff}$ and $\omega^{DE}$ in time in accordance with the energy conditions outcomes. In Section 5 we further discuss our results and the matter-geometry coupling issue. We also propose other areas in which the $f(R,T)$ matter-geometry coupling consideration may generate interesting and testable outcomes.

\section{The $f(R,T)=f_1(R)+f_2(R)f_3(T)$ gravity}\label{sec:frt}

The total action in the $f(R,T)$ theory of gravity reads \cite{harko/2011}

\begin{equation}\label{frt1}
S=\frac{1}{16\pi}\int d^{4}x\sqrt{-g}f(R,T)+\int d^{4}x\sqrt{-g}L_m,
\end{equation}
with $g$ being the metric determinant.

By varying this action with respect to the metric yields

\begin{align}\label{frt2}
[f_1'(R)+f_2'(R)f_3(T)]R_{\mu\nu}-\frac{1}{2}f_1(R)g_{\mu\nu}+  \\ \nonumber
(g_{\mu\nu}\Box-\nabla_\mu\nabla_\nu)[f_1'(R)+f_2'(R)f_3(T)]=[8\pi+ \\ \nonumber
f_2(R)f_3'(T)]T_{\mu\nu}+ f_2(R)\left[f_3'(T)p+\frac{1}{2}f_3(T)\right]g_{\mu\nu},
\end{align}
for which it was assumed $f(R,T)=f_1(R)+f_2(R)f_3(T)$ and primes denote derivatives with respect to the argument.

Now, we will take $f_1(R)=f_2(R)=R$ and $f_3(T)=\alpha T$, with $\alpha$ a constant. This is the simplest non-trivial functional form of the function $f(R,T)$ which involves non-minimal matter-geometry coupling within the $f(R,T)$ formalism. Moreover, it benefits from the fact that GR is retrieved when $\alpha=0$.

The considerations above yield, for Eq.(2), the following
 
\begin{equation}\label{frt3}
G_{\mu\nu}=8\pi T_{\mu\nu}^{eff}=8\pi(T_{\mu\nu}+T_{\mu\nu}^{DE}),
\end{equation}
with $T_{\mu\nu}^{eff}$ being the effective energy-momentum tensor, $T_{\mu\nu}$ the usual matter energy-momentum tensor and the dark energy term $T_{\mu\nu}^{DE}$, coming from the matter-geometry coupling predicted in the present theory, is written as

\begin{equation}\label{frt3.1}
T_{\mu\nu}^{DE}=\frac{\alpha R}{8\pi}\left(T_{\mu\nu}+\frac{3\rho-7p}{2}g_{\mu\nu}\right),
\end{equation} 
in which the coupling terms can be straightforwardly noticed.

By applying the Bianchi identities in Equation (\ref{frt3}) yields

\begin{equation}\label{frt4}
\nabla^{\mu}T_{\mu\nu}=-\frac{\alpha R}{8\pi}\left[\nabla^{\mu}(T_{\mu\nu}+pg_{\mu\nu})+\frac{1}{2}g_{\mu\nu}\nabla^{\mu}(\rho-3p)\right].
\end{equation}
As required, note that by taking $\alpha=0$ in Eqs.(3)-(5) retrieves GR formalism.

\section{The $f(R,T)=R+\alpha RT$ cosmology}

For a flat Friedmann-Robertson-Walker universe with scale factor $a(t)$ and Hubble parameter $H=\dot{a}/a$, the non-null components of (\ref{frt3}), for $\rho^{eff}=\rho+\rho^{DE}$ and $p^{eff}=p+p^{DE}$, are

\begin{equation}\label{frt5}
3H^2=8\pi\rho^{eff},
\end{equation}
\begin{equation}\label{frt6}
2\dot{H}+3H^2=-8\pi p^{eff},
\end{equation}
with dots being time derivatives and

\begin{equation}\label{frt5.1}
\rho^{eff}=\rho-\frac{3\alpha}{8\pi}\left(\dot{H}+2H^2\right)(3\rho-7p), 
\end{equation}
\begin{equation}
p^{eff}=p+\frac{9\alpha}{8\pi}\left(\dot{H}+2H^2\right)(\rho-3p).
\end{equation}
Moreover, Eq.(\ref{frt4}) reads

\begin{equation}\label{frt7}
\dot{\rho}+3H(\rho+p)=\left[1-\frac{4\pi}{\frac{3}{\alpha}\left(\dot{H}+2H^2\right)}\right]^{-1}(\dot{p}-\dot{\rho}).
\end{equation}

From Equations (\ref{frt5}) and (\ref{frt6}), we have

\begin{equation}
\rho=\frac{H^2\left[8\pi-27\alpha\left(\dot{H}+2H^2\right)\right]+7\alpha(2\dot{H}+3H^2)\left(\dot{H}+2H^2\right)}{\frac{64\pi^2}{3}-96\pi \alpha \left(\dot{H}+2H^2\right)+18\alpha^2 \left(\dot{H}+2H^2\right)^2}, \hspace{1.0cm}
\end{equation}

\begin{equation}
p=-\frac{9\alpha H^2\left(\dot{H}+2H^2\right)+\left(2\dot{H}+3H^2\right)\left[\frac{8\pi}{3}-3\alpha\left(\dot{H}+2H^2\right)\right]}{\frac{64\pi^2}{3}-96\pi \alpha \left(\dot{H}+2H^2\right)+18\alpha^2 \left(\dot{H}+2H^2\right)^2}.
\end{equation}

In order to find the solutions for $\rho(t)$ and $p(t)$ we need to know $H(t)$. A great number of parametrization schemes have been investigated in the literature with the requirement of their theoretical consistency and observational viability. In particular, we can quote the power-law expansion $(a\propto t^n)$ and exponential law $(a\propto e^{mt})$, with $n$ and $m$ being non-negative constants. 

Here we consider a simple ansatz which is obtained by multiplying the power and exponential laws, called hybrid expansion law (HEL). Such an ansatz mimics the power-law and de Sitter cosmologies as special cases, but, as it will be shown below, it also provides an elegant description of the transition from decelerated to accelerated cosmic expansion.

It has been tested from observational data referred to Big Bang Nucleosynthesis, Baryon Acoustic Oscillations and Cosmic Microwave Background  \cite{Akarsu/2014}. The authors in \cite{Akarsu/2014} have shown that all the cosmological parameters related with the present day universe as well as with the onset of the cosmic acceleration for HEL and $\Lambda$CDM models are consistent within the $1\sigma$ confidence level. They also gave the values of some important cosmological parameters with $1\sigma$ errors for both models at early and future epochs, showing that they exhibit similar behaviors at future epochs.

We consider, then, as a solution for the scale factor, the HEL in the form

\begin{equation}
a(t)=e^{mt}t^n,
\end{equation}
so that the Hubble and deceleration parameters are 

\begin{equation}
H=m+\frac{n}{t},
\end{equation}
\begin{equation}
q=-\frac{\ddot{a}}{aH^2}=-1+\frac{n}{(mt+n)^2}.
\end{equation}
Here, one can choose the constants in such a way that the power-law dominates over exponential law in the early universe and 
the exponential law dominates over power-law at late times, in order to account for the present acceleration of the universe expansion \cite{riess/1998,perlmutter/1999}.

From Equation (15) it is clear that there is a transition phase from deceleration to acceleration at $t=-\frac{n}{m}\pm\frac{\sqrt{n}}{m}$ with $0<n<1$. Since the negativity of the second term leads to a negative time, which indicates an unphysical context of the Big Bang cosmology, we conclude that the cosmic transition may have occurred at $t=\frac{\sqrt{n}-n}{m}$.

Figure 1 below presents the deceleration parameter evolution in time, obtained above for the hybrid scale factor. 

\begin{figure}[ht]
\centering
\includegraphics[width=0.5\textwidth]{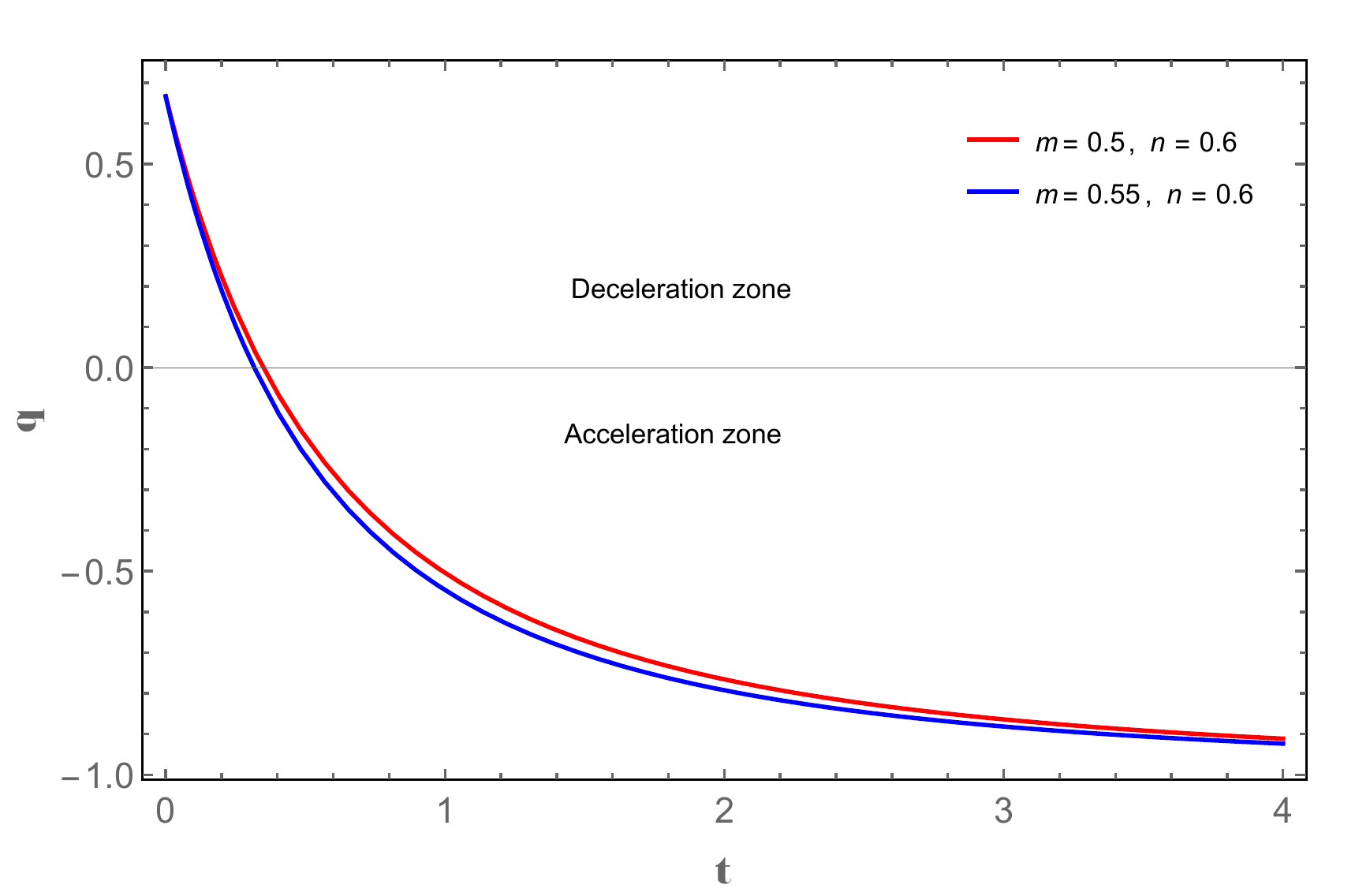}
\caption{Deceleration parameter evolution in time.}
\end{figure}

From $a(t)=\frac{1}{1+z}$, with $z$ being the redshift and the present scale factor $a_0=1$, we obtain the following time-redshift relation

\begin{equation}
t=\frac{n W\left[\frac{m \left(\frac{1}{z+1}\right)^{1/n}}{n}\right]}{m},
\end{equation}
where $W$ denotes the Lambert function (also known as ``product logarithm"). 

By using Equation (16), we can plot the deceleration parameter with respect to the redshift, which can be appreciated in Fig.2 below.

\begin{figure}[ht]
\centering
\includegraphics[width=0.5\textwidth]{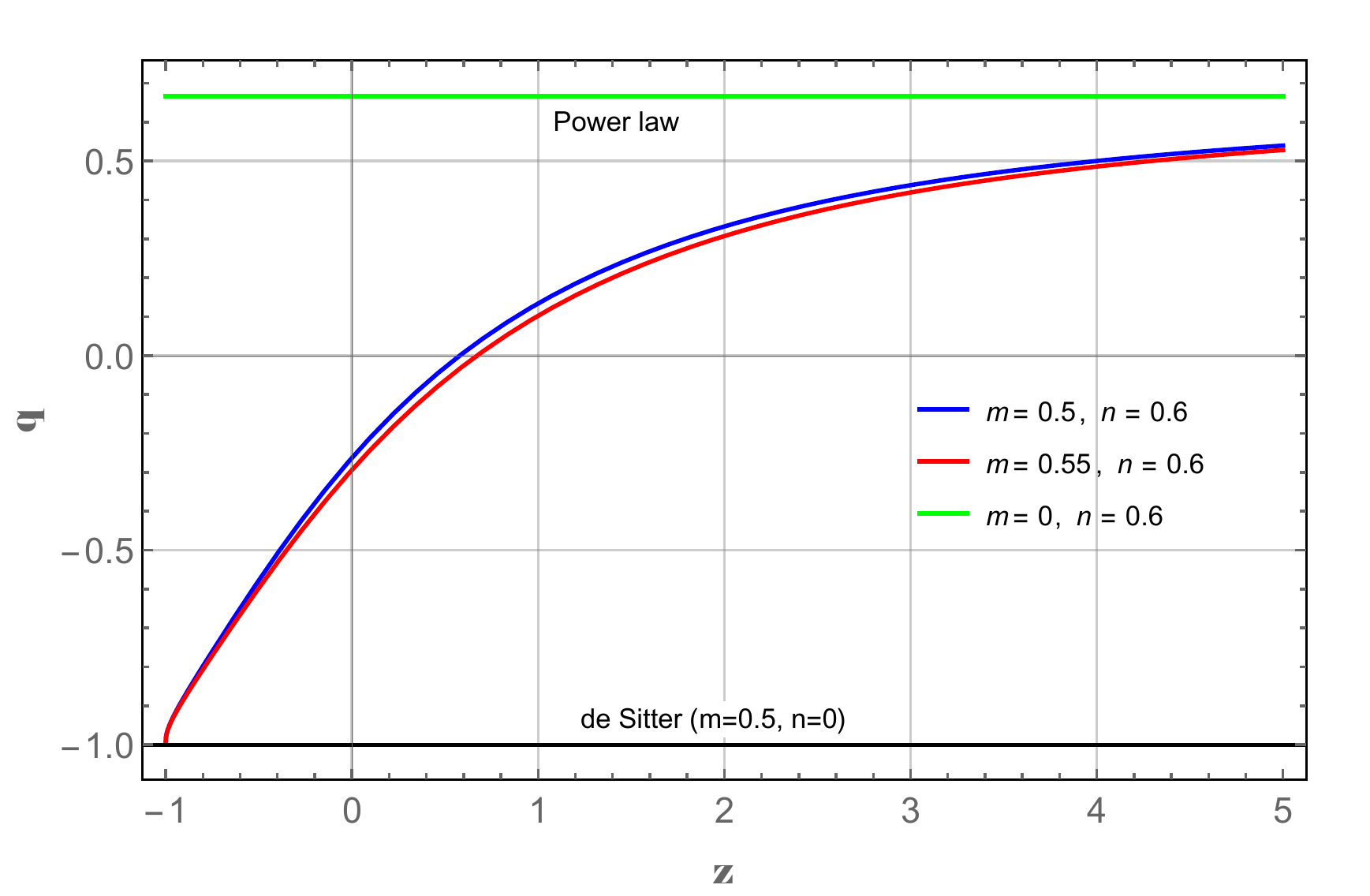}
\caption{Deceleration parameter evolution in redshift.}
\end{figure}

The transition from the decelerated to the accelerated phase of the universe expansion occurs at some redshift, which we will call $z_{tr}$ and in our model it depends directly on the parameter $m$. From Figure 2, such a transition occurs at $z_{tr} =0.578825,0.671744$, corresponding to $m=0.5,0.55$, respectively. These values are in accordance with recent observational data \cite{Capozziello14,Capozziello15,Farooq17}.

\section{Energy conditions application and its consequences on cosmological quantities}

The energy conditions are based on the Raychaudhuri equation, which describes the behaviour
of the compatibility of timelike, lightlike or spacelike curves. It is commonly used in GR to establish and study the singularities of spacetime \cite{Wald84}. In particular, in \cite{Alvarenga/2012}, F.G. Alvarenga et al. have tested the energy conditions in $f(R,T)$ theory of gravity. 

In this section we will apply the energy conditions to our solutions for the effective energy density and effective pressure.

The well known point-wise energy conditions are the following:

\begin{itemize}
\item Strong energy condition (SEC): gravity should be always attractive, and in cosmology the relation $\rho^{eff}+3p^{eff}\geqslant 0$ must be obeyed;
\item Weak energy condition (WEC): the effective energy density should always be non-negative when measured by any observer, i.e., $\rho^{eff}>0, \ \ \rho^{eff}+p^{eff}\geqslant 0$;
\item Null energy condition (NEC): it is the minimum requirement which is obtained from SEC and WEC, i.e., $\rho^{eff}+p^{eff}\geqslant 0$;
\item Dominant energy condition (DEC): the effective energy density must always be positive when measured by any observer, i.e., the relation $\rho^{eff} \geqslant \vert p^{eff} \vert$ must be obeyed.
\end{itemize}

In order to obtain the equations for $\rho^{eff}$ and $p^{eff}$ as functions of $t$, we firstly substitute the solution (14) for $H$ in Equations (11)-(12), which makes us able to write the ordinary matter energy density $\rho$ and pressure $p$ as Equations (17)-(18) below:

\newpage

\begin{widetext}
\begin{equation}
\rho=\frac{12 \pi  t^2 (m t+n)^2-3 \alpha  \left[2 m^2 t^2+4 m n t+n (2 n-1)\right] \left[3 m^2 t^2+6 m n t+n (3 n+7)\right]}{27 \alpha ^2 \left[2 m^2 t^2+4 m n t+n (2 n-1)\right]^2-144 \pi  \alpha  t^2 \left[2 m^2 t^2+4 m n t+n (2 n-1)\right]+32 \pi ^2 t^4},
\end{equation}
\begin{equation}
p=-\frac{2 n t^2 \left[9 \alpha  m^2+4 \pi  (3 m t-1)\right]+12 \pi  m^2 t^4+3 n^2 \left[3 \alpha  (4 m t-1)+4 \pi  t^2\right]+18 \alpha  n^3}{27 \alpha ^2 \left[2 m^2 t^2+n (4 m t-1)+2 n^2\right]^2-144 \pi  \alpha  t^2 \left[2 m^2 t^2+n (4 m t-1)+2 n^2\right]+32 \pi ^2 t^4}.
\end{equation}
\end{widetext}

Now, by putting Equations (17)-(18) in (8)-(9), we obtain the following expressions for $\rho^{eff}$ and $p^{eff}$: 
\begin{equation}
\rho^{eff}=\frac{3 (m t+n)^2}{8 \pi  t^2},
\end{equation}
\begin{equation}
p^{eff}=\frac{-3 m^2 t^2-6 m n t-3 n^2+2 n}{8 \pi  t^2}.
\end{equation}

We can understand the energy conditions as able to provide us the validity regions of our solutions, since they evade, for instance, the presence of space-time singularities. 

From (19)-(20), we become able to plot the energy conditions as Figures 3-5 for fixed (observationally validated) value of $n$.

\begin{figure}
\centering
\includegraphics[width=0.38\textwidth]{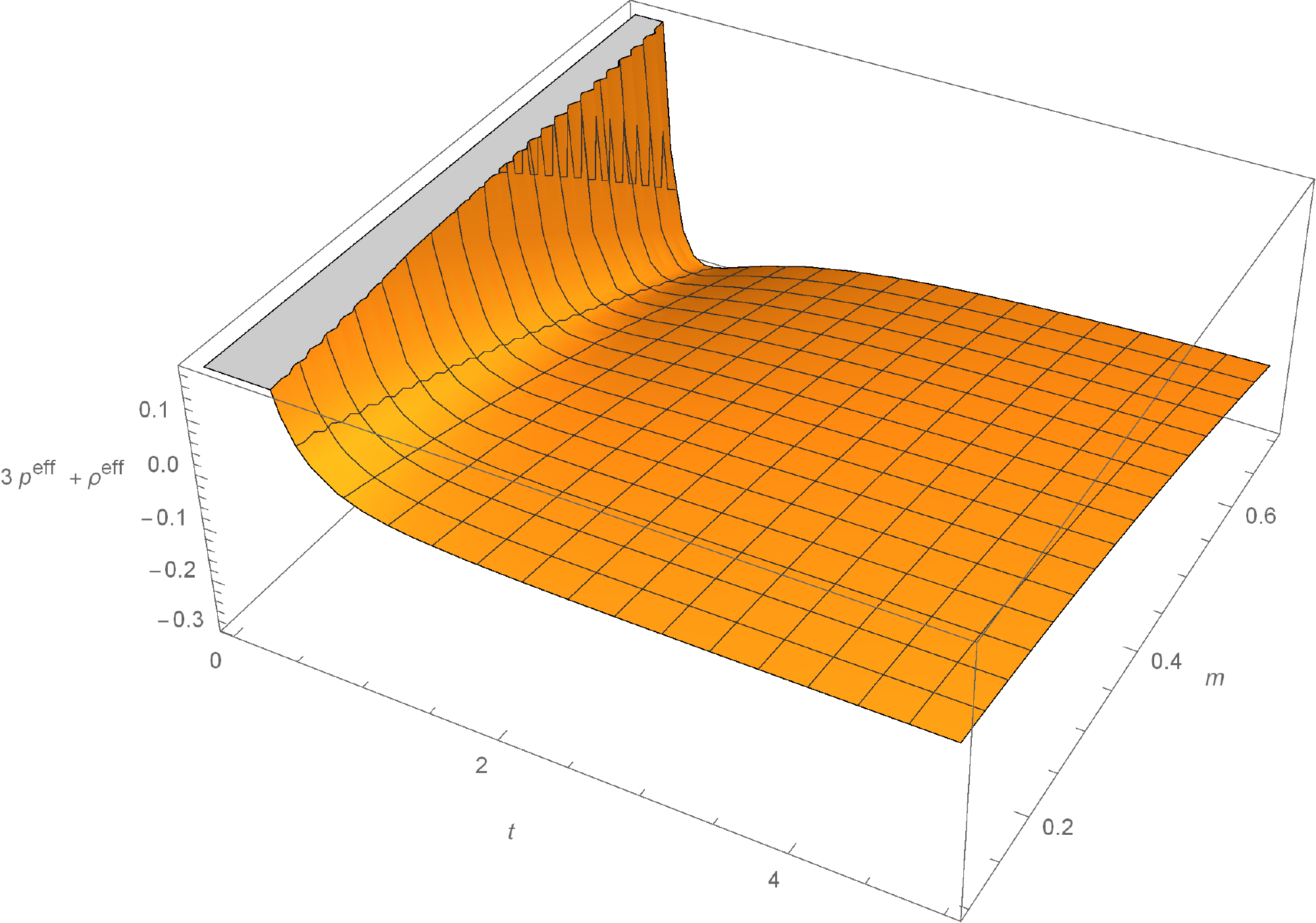}
\caption{$\rho^{eff}+3p^{eff}$ vs. $t$, with $n=0.6$.}
\end{figure}

\begin{figure}
\centering
\includegraphics[width=0.38\textwidth]{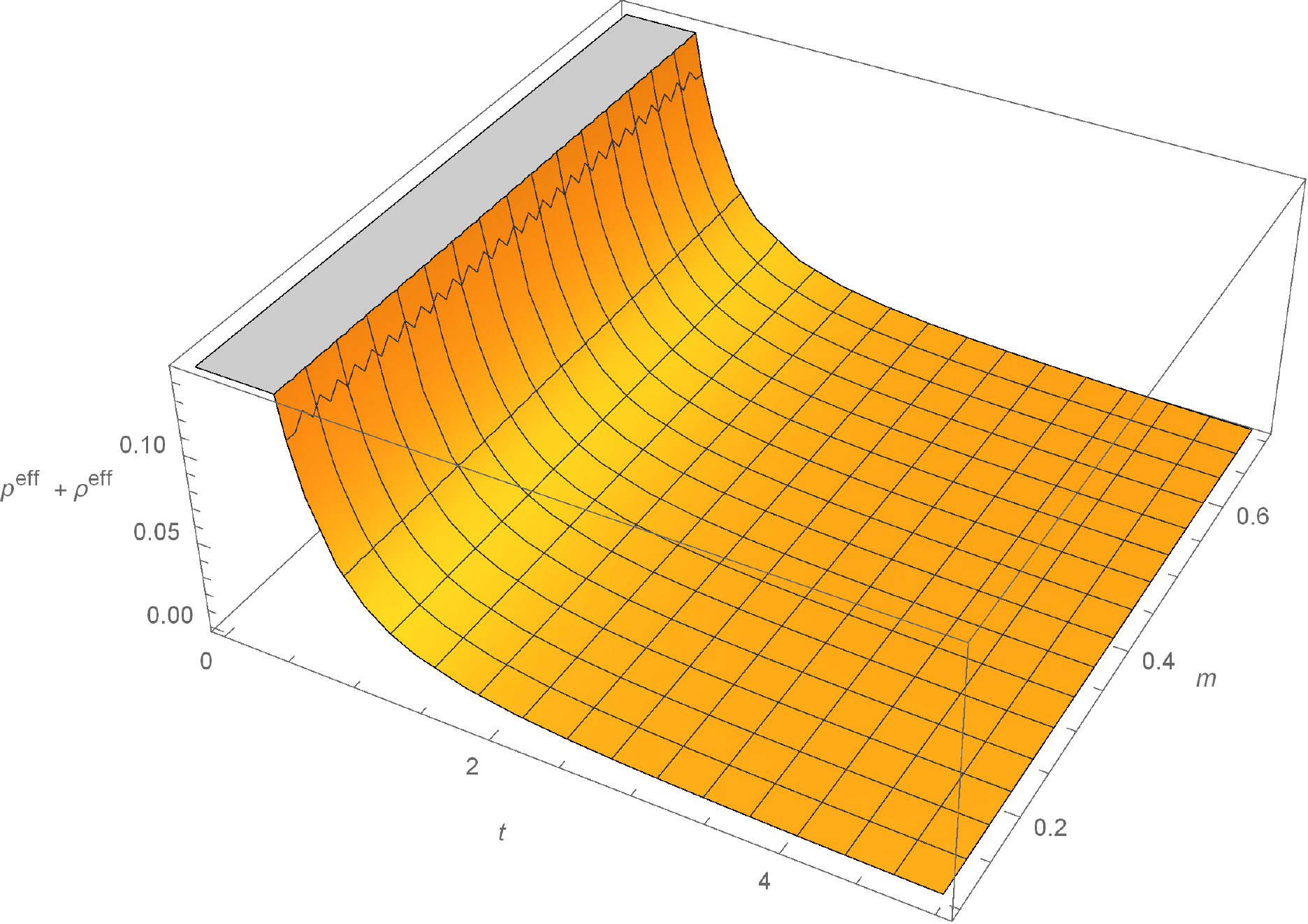}
\caption{$\rho^{eff}+p^{eff}$ vs. $t$, with $n=0.6$.}
\end{figure}

\begin{figure}
\centering
\includegraphics[width=0.38\textwidth]{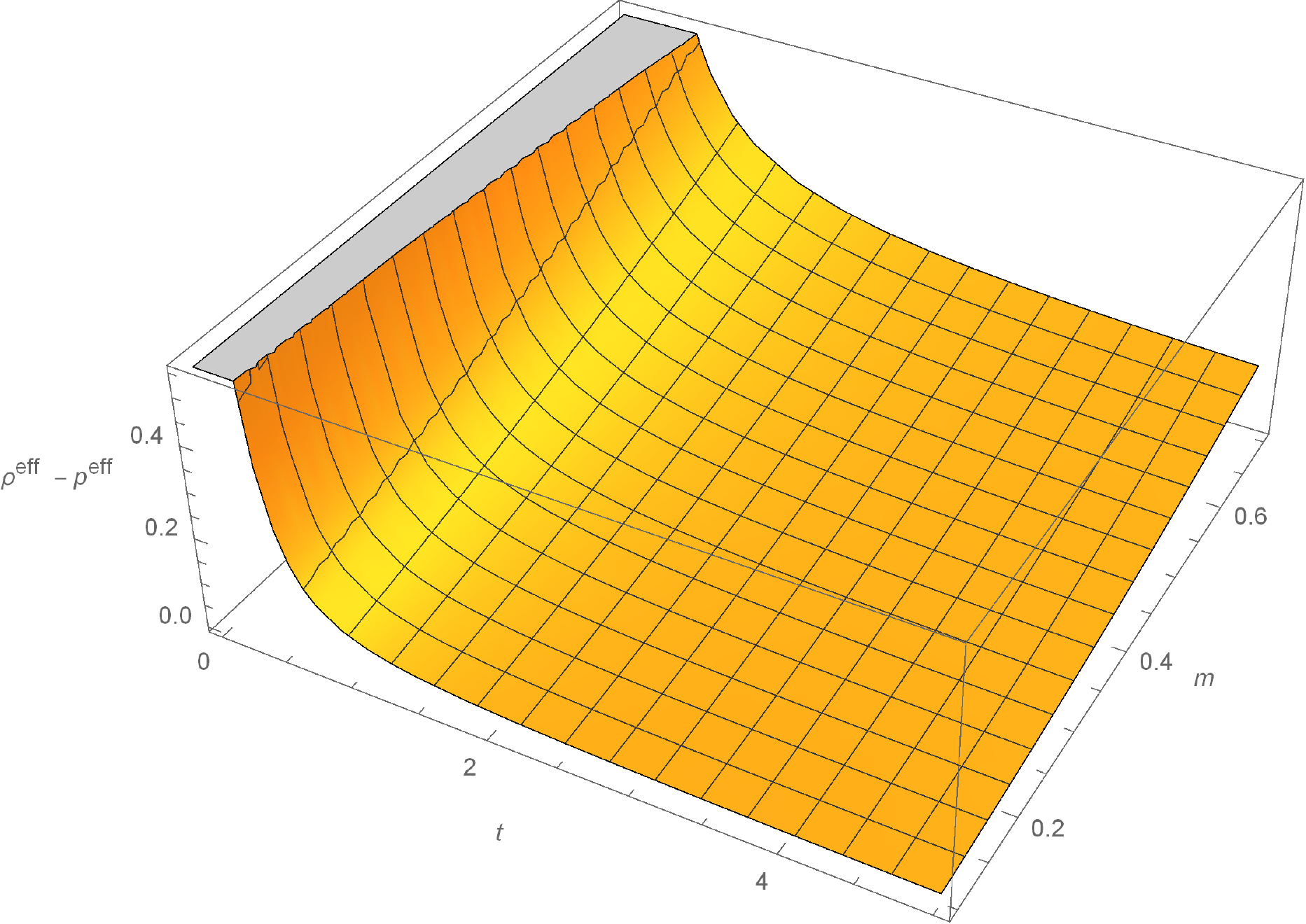}
\caption{$\rho^{eff}-p^{eff}$ vs. $t$, with $n=0.6$.}
\end{figure}

The evolution of effective energy density, pressure and EoS in time are given in Figures 6, 7 and 8.

\begin{figure}
\centering
\includegraphics[width=0.38\textwidth]{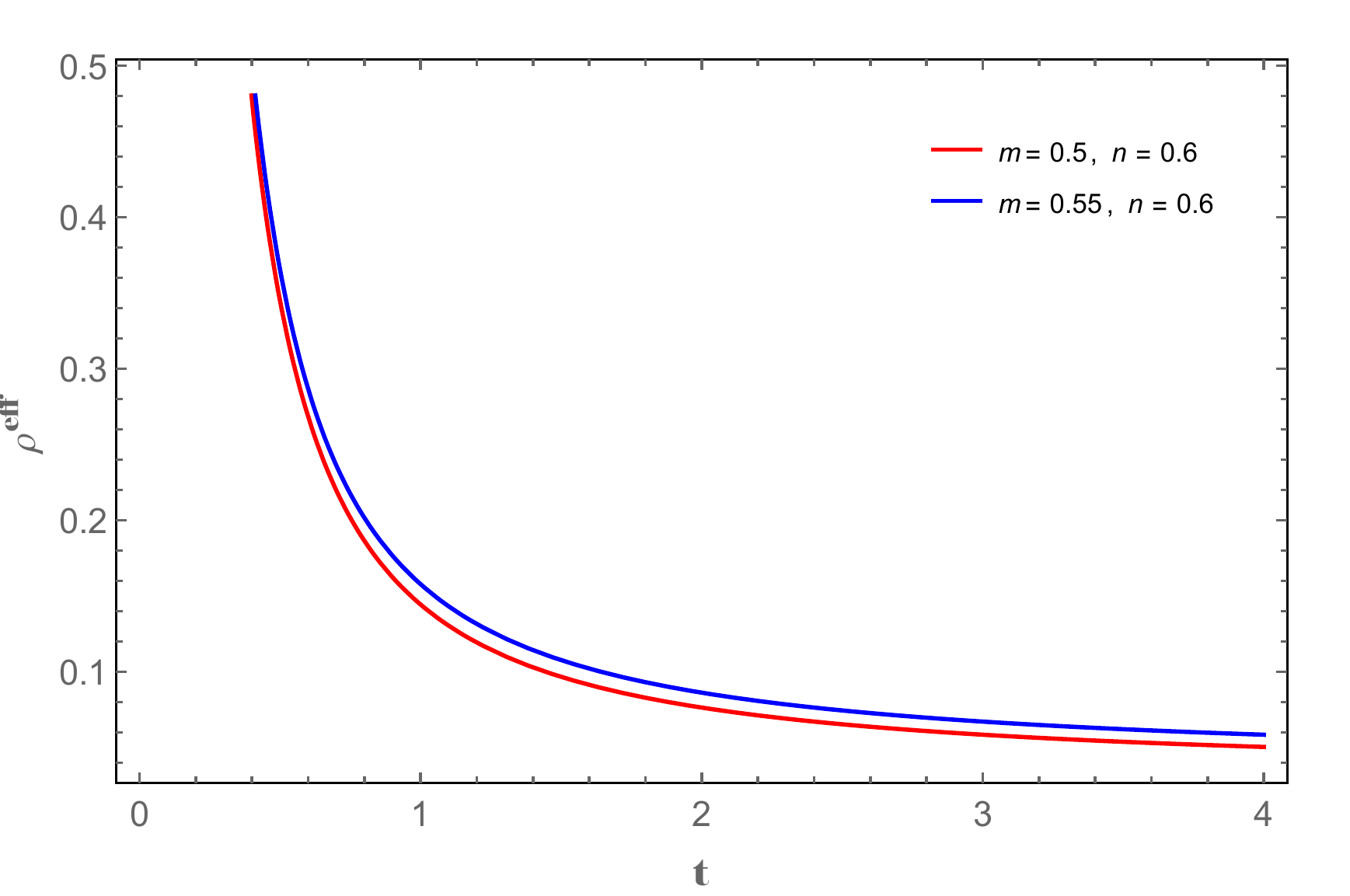}
\caption{Evolution of $\rho^{eff}$ in time.}
\end{figure}

\begin{figure}
\centering
\includegraphics[width=0.38\textwidth]{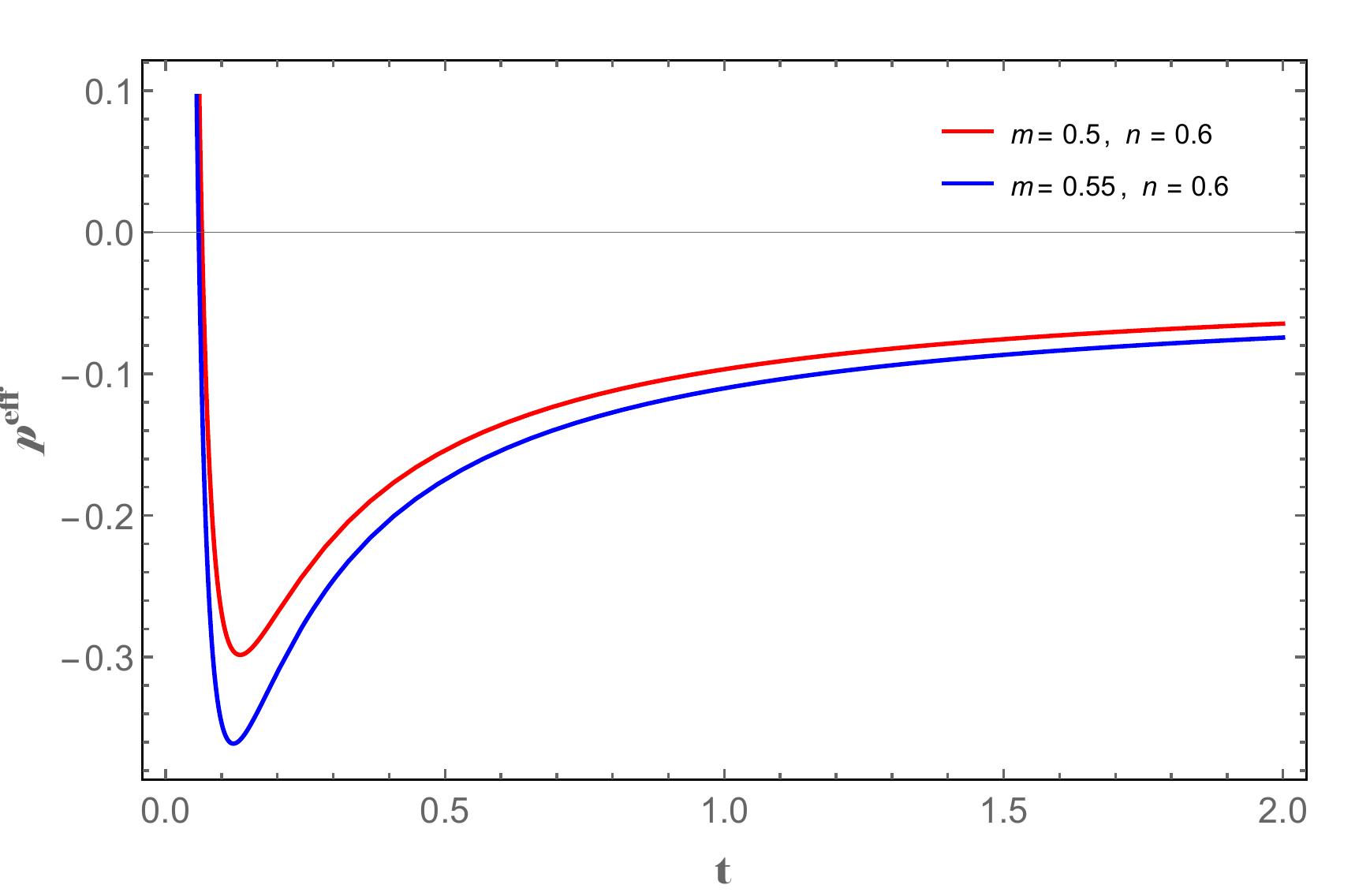}
\caption{Evolution of $p^{eff}$ in time.}
\end{figure}

\newpage

\begin{figure}
\centering
\includegraphics[width=0.38\textwidth]{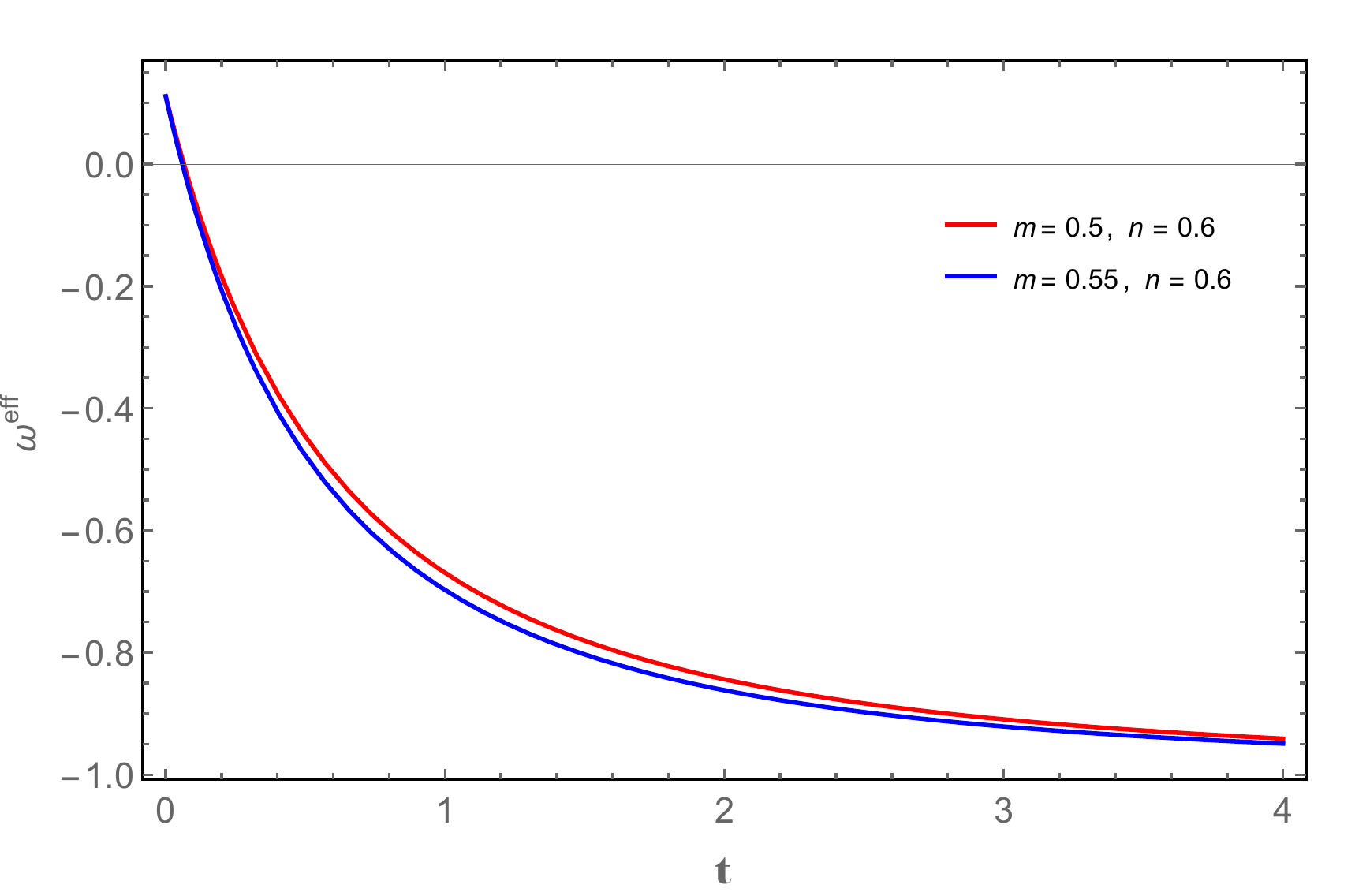}
\caption{Evolution of $\omega^{eff}$ in time.}
\end{figure}

Now we can also write the asymptotic behaviour of the cosmological parameters $a$, $q$, $\rho^{eff}$ and $p^{eff}$, which is presented in Table 1.

\begin{table}[ht]
\begin{tabular}{|c|c|c|}
  \hline
 Parameters & $t\rightarrow 0 (z\rightarrow \infty)$ & $t\rightarrow \infty (z\rightarrow -1)$ \\\hline
 $a$ & 0 & $\infty$ \\
 $q$ &$\frac{1}{n}-1$ & -1  \\  
 $\rho^{eff}$ & $\infty$ & $0$  \\
 $p^{eff}$ & $ \infty$ & $0$  \\
     \hline
\end{tabular}
\caption{Asymptotic behaviour of the cosmological parameters.}
\end{table}
In this way, the dark energy contribution for energy density and pressure, according to the model, read
\begin{widetext}
\begin{equation}
\rho^{DE}=\frac{3 (m t+n)^2}{8 \pi  t^2}-\frac{12 \pi  t^2 (m t+n)^2-3 \alpha  \left[2 m^2 t^2+4 m n t+n (2 n-1)\right] \left[3 m^2 t^2+6 m n t+n (3 n+7)\right]}{27 \alpha ^2 \left[2 m^2 t^2+4 m n t+n (2 n-1)\right]^2-144 \pi  \alpha  t^2 \left[2 m^2 t^2+4 m n t+n (2 n-1)\right]+32 \pi ^2 t^4},
\end{equation}
\begin{equation}
p^{DE}=-\frac{27 \alpha  \left[2 m^2 t^2+4 m n t+n (2 n-1)\right]^2 \{t [3 \alpha  m (m t+2 n)-8 \pi  t]+\alpha  n (3 n-2)\}}{8 \pi  t^2 \left\{27 \alpha ^2 \left[2 m^2 t^2+4 m n t+n (2 n-1)\right]^2-144 \pi  \alpha  t^2 \left[2 m^2 t^2+4 m n t+n (2 n-1)\right]+32 \pi ^2 t^4\right\}}.
\end{equation}
\end{widetext}

The dark energy EoS parameter has the analytical form presented in Eq.(23): 

\begin{widetext}
\begin{equation}
\omega^{DE}=-\frac{9 \left[(2 m^2 t^2+4 m n t+n (2 n-1)\right] \{t [3 \alpha  m (m t+2 n)-8 \pi  t]+\alpha  n (3 n-2)\}}{27 \alpha  (m t+n)^2 \left[2 m^2 t^2+4 m n t+n (2 n-1)\right]-8 \pi  t^2 \left[15 m^2 t^2+30 m n t+n (15 n-7)\right]}
\end{equation}
\end{widetext}

When plotting such a parameter, it is vital to respect the range of acceptable values for $\alpha$, according to the energy conditions presented above. Note that such a consideration was not applied to the deceleration parameter since this quantity does not depend on the value of $\alpha$. $\omega^{DE}$ is plotted against time in Figure 9 below.

\begin{figure}
\centering
\includegraphics[width=0.4\textwidth]{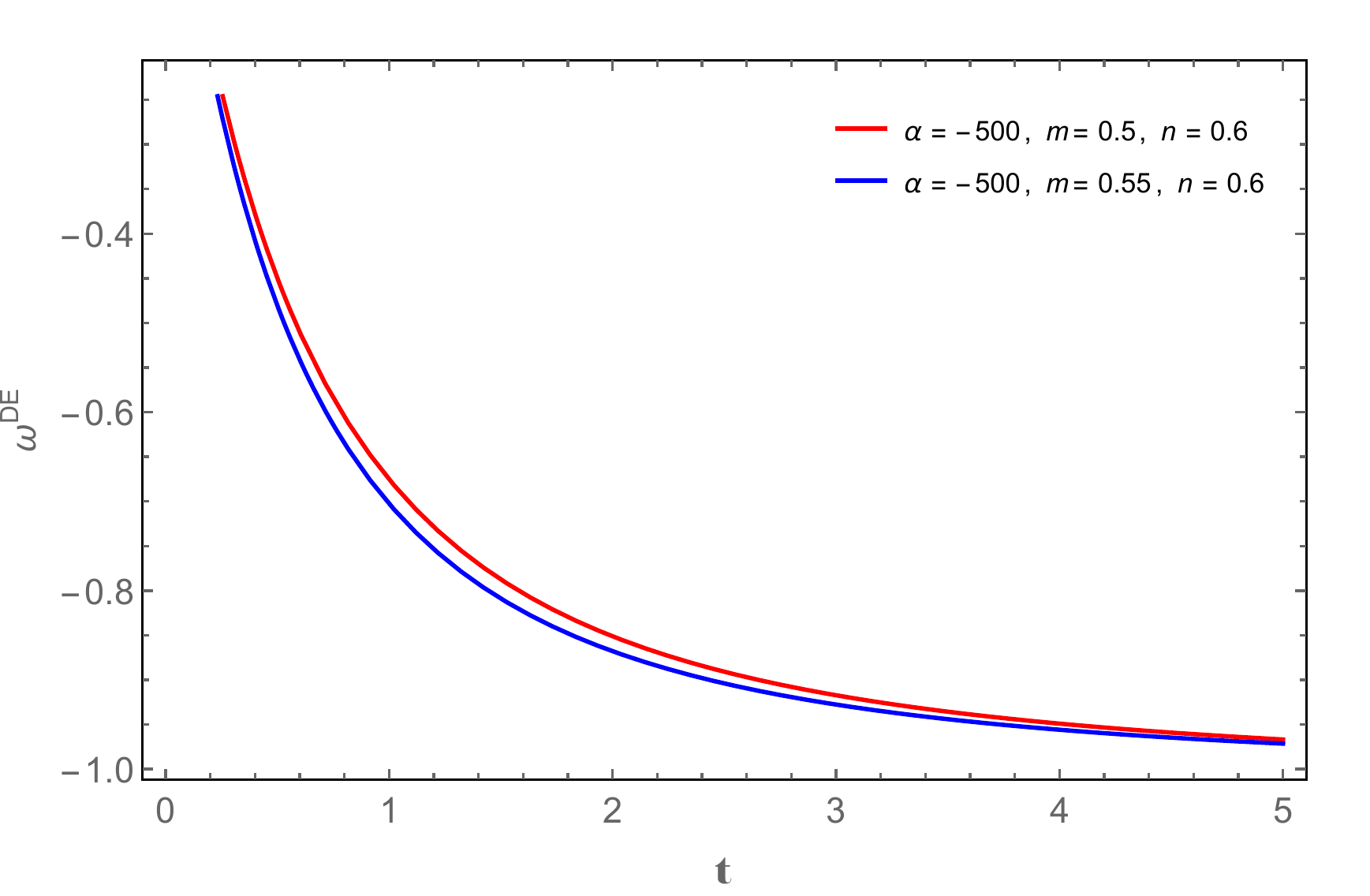}
\caption{Evolution of $\omega^{DE}$ in time.}
\end{figure}

Some interesting and relevant cosmological features are present in Figures 7-9, as it will be discussed in the next section.

\section{Discussion}

In this article, we have constructed, as a pioneer proposal, a cosmological scenario from the simplest non-minimal matter-geometry coupling in the $f(R,T)$ gravitational theory. In the present section, we will discuss the energy conditions applications and the cosmological viability of the model. Moreover, we argue about the matter-geometry coupling issue, which is a model premise.

In what concerns the results obtained from the energy conditions application in Section 4, it is worth stressing that due to the current accelerated expansion of the universe \cite{riess/1998,perlmutter/1999}, SEC must be abandoned \cite{VISSER/2000,VISSER/2002}. This is a consequence of the fact that from standard Friedmann equations, an accelerated expansion universe should be driven by an exotic fluid of EoS parameter $<-1/3$. From Figure 3, we see that the SEC is indeed violated in our model. One can observe that minimally coupled and curvature coupled scalar field theories also violate SEC \cite{VISSER/2002}. 

On the other hand, the WEC, NEC and DEC are satisfied in our model as it can be checked in Figures 4-5.   

Let us now check the cosmological viability of our model. In Section 3 we showed that the deceleration parameter respects the observational constraints and predicts a transition from a phase of deceleration to a phase of acceleration of the universe expansion. Such a transition occurs in a redshift $z_{tr}$ which agrees with recent observational data.

We shall highlight that the transition phenomenon can also be noticed in the evolution of the effective pressure in time, as Fig.7. In such a figure, both curves predict the pressure of the universe to eventually assume negative values. It is well known that a negative pressure fluid is the exact mechanism able to explain a cosmic acceleration within standard cosmology, although in the latter it is necessary to invoke the cosmological constant in order to obtain such an exotic feature. 

The effective EoS parameter in Fig.8 also presents some properties for which a more profound discussion is worthy. Firstly, one should note that, once again, the decelerated-accelerated expansion of the universe transition is being predicted. As time passes by, $\omega^{eff}$ decreases its value and the region in which $\omega^{eff}<-1/3$ represents an epoch of cosmic acceleration, according to Friedmann equations.

For high values of $t$, $\omega^{eff}\rightarrow-1$, which is the current value of the EoS parameter of the universe according to observations of fluctuations on the cosmic microwave background temperature \cite{hinshaw/2013}.

If we analyse only the coupled $f(R,T)$ contribution to the EoS parameter in Fig.9, we see that $\omega^{DE}$ values are restricted to the range $\omega^{DE}<-1/3.$ As mentioned above, this range for the values of an EoS is capable of inducing the effects of a cosmic acceleration. 

The cosmic acceleration has already been considered within matter-geometry coupling models predicted from an extension of the $f(R)$ formalism which presents a term like $f(R)L_m$ in the action \cite{allemandi/2005,bertolami/2010,bisabr/2012}. Departing from the usual $f(R)$ formalism, these models do not have a divergence free energy-momentum tensor and this is a consequence of the transferring energy and momentum between matter and geometry.

In the present model, we can interpret the prediction of cosmic acceleration as a consequence of energy transference between geometry and matter. In other words, such a transferring process is able to provide an effective fluid of sufficient negative pressure, responsible for driving the cosmic speed up.

The second term on the {\it rhs} of Eq.(3), which is the responsible for the non-conservation of the matter energy-momentum tensor, induces the movement of test particles in the presence of a gravitational field to be non-geodesic \cite{harko/2011}. O. Bertolami and J. P\'aramos showed that the discrepancy between classical theoretical prediction of galactic rotation curves and flatness observations may be due to deviation from geodesic motion \cite{bertolami/2010c}. In this way, the implications of $\nabla_\mu T^{\mu\nu}\neq0$ might be observed in galactic rotation curves.

The consequences of matter-geometry coupling can also appear in other applications of $f(R,T)=R+\alpha RT$ gravity. For instance, the analysis of gravitational waves in such a model may yield the propagation velocity of gravitational waves to be $\neq c$ \cite{amam/2016}. The dark matter should also be investigated within this formalism, as the non-minimal coupling has already been used to mimic its effects in other gravitational models \cite{bertolami/2010b}.

\begin{acknowledgements}
PHRSM would like to thank S\~ao Paulo Research Foundation (FAPESP), grant 2015/08476-0, for financial support. PKS acknowledges the support of CERN, in Geneva, during an academic visit, where a part of this work was done. The authors also thank the referee for the valuable suggestions which improved the presentation of the formalism as well as of the results obtained.
\end{acknowledgements}



\end{document}